\documentclass [12pt,a4paper]{article}
\begin{document}
\parindent=0cm
\parskip=1.5mm
\def\bi{\begin{list}{$\bullet$}{\parsep=0.5\baselineskip
      \topsep=\parsep \itemsep=0pt}} \def\ei{\end{list}}
\def\phi{\varphi} \def\-{{\bf --}} \def\vm{v_{max}}
\newcommand{\eins}{\mathbf 1}
\begin{center}
  {\LARGE\bf Exact results for one dimensional}
\end{center}
\begin{center}
  {\LARGE\bf stochastic cellular automata}
\end{center}
\begin{center}
  {\LARGE\bf for different types of updates}
\end{center}
\vskip1.8cm \setcounter{footnote}{0}
\begin{center}
  {\Large N.\ Rajewsky\footnote{email: \tt nr@thp.uni-koeln.de} 
    and M.\ 
    Schreckenberg\footnote{email: \tt schreck@math.uni-duisburg.de} }
\end{center}
\vskip1.3cm
\begin{center}
  $^{1}$\,Institut f\"ur Theoretische Physik\\ Universit\"at zu K\"oln\\ 
  D--50937 K\"oln, F.R.G.
\end{center}
\begin{center}
  $^{2}$ FB 11--Mathematik\\ Gerhard-Mercator-Universit\"at
  Duisburg\\ D--47048 Duisburg, F.R.G.\\ 
\end{center}
\vskip1.3cm \vskip2cm {\large \bf Abstract}\hspace{.4cm} We study two
common types of time-noncontinuous updates for one dimensional
stochastic cellular automata with arbitrary nearest neighbor
interactions and arbitrary open boundary conditions.  We first
construct the stationary states using the matrix product formalism.
This construction then allows to prove a general connection between
the stationary states which are produced by the two different types of
updates. \\
Using this connection, we derive explicit relations between the
densities and correlation functions for these different 
stationary states. 
 \vfill \pagebreak In this letter we study one dimensional
stochastic cellular automata with different types of update.  We
consider a chain of sites $i=1,...,L$ ($L$ even) and assign a discrete
variable $\tau_i=1,...,m$ to each site $i$. $\tau_i$ can for example
represent $m$ different sorts of particles. We then prescribe some
local rule ${\cal T}$ describing the interaction between a pair of
nearest neighbor variables ($\tau_i,\tau_{i+1}$) and, as usually done,
choose this rule to be independent of the location of the sites. One
can further allow for some rules ${\cal R}$ and ${\cal{L}}$, acting on
the sites at the two ends of the chains. Such boundary conditions
could for example mimic the injection and removal of particles with
certain rates. This
type of boundaries can lead to a rich dynamical behaviour of the
model. One prominent example is the asymmetric exclusion
model (see \cite{zia}, \cite{ligget} and ref. therein)
which exhibits boundary induced phase
transitions where the bulk density of particles depends in a subtle
way on the boundary conditions. \\ There are now several ways to
implement our prescribed update rules. These implementations are an
essential part of the definition of a model since they generally
produce different transients, stationary states and averages of
observables.\\ For analytic calculations one usually uses the {\it
  random sequential} update. One picks at random site $i$ and applies
the update rule. This update generally produces the weakest
correlations between the sites compared to other update procedures.
The master equation which describes the temporal evolution of the
weights of the configurations $\{\tau_1,\tau_2,...,\tau_L\}$ can then
be written as a Schr\"odinger-like equation with a (non--hermetian)
hamiltonian consisting of a sum of local hamiltonians. For this update
it was recently shown, that the stationary state can be always written
as the scalar product of the product of some (generally) noncommuting
matrices \cite{sandow96}.\\ There are three other basic ways of
implementing the update rules which are much more common for practical
purposes like computer simulations, since the random sequential update
requires an additional random number for each local update. These
three types are called {\it sequential}, {\it sublattice parallel} and
{\it parallel} update.  In the parallel update one applies the update
rules simultaneously to all sites. 
This case will not be discussed in the following. Note that the parallel
update usually produces the strongest correlations between the sites
and is often used in the context of traffic flow models {\cite{traffic}}.\\
In the {sequential} update we start at the right
end of the chain and apply our boundary condition ${\cal R}$ to site
$i=L$. We then update the pair ($\tau_{L-1},\tau_{L}$). After that we
proceed with the pair ($\tau_{L-2},\tau_{L-1}$) and so forth until we
reach the left end of the chain where we apply ${\cal L}$ at site
$i=1$. This procedure can of course also be done from the left to the
right. Let us denote these two updates with $T_{\rightarrow}$ and
$T_{\leftarrow}$.\\ The {sublattice parallel} update $T_{\parallel}$
is defined as follows: In the first time step one applies ${\cal L}$,
${\cal R}$ and updates all pairs ($\tau_i,\tau_{i+1}$) with an even
$i$.  In the second time step one updates all pairs
($\tau_i,\tau_{i+1}$) with an odd $i$.\\ In the following, we show
that in the stationary state averages like densities and correlation
functions of the {sequential} and {sublattice parallel} update are
essentially the same. This is a nontrivial result since the
microscopic configurations of the system are generally very different
after applying a {sequential} or a {sublattice parallel} update. This
result should be of practical use for people doing computer
simulations of cellular automata or reaction-diffusion processes. Note
that the {sequential} update can transport a particle many sites
during one update of the whole chain while $T_{\parallel}$ can move a
particle only two sites at maximum.\\ We start with the construction
of the stationary state of an arbitrary model with {\it sequential }
update $T_{\leftarrow}$ from the right to the left end. We will first
show that it is always possible to write the stationary probability
distribution $P_0(\tau_1,\tau_2,...,\tau_L)$ of the configurations
$\{{\tau_i}\}$ as
\begin{equation}\label{produkt}
  P_0(\tau_1,\tau_2,...,\tau_L)_{\leftarrow}=
\langle W|A_{\tau_1}A_{\tau_2}...A_{\tau_L}|V\rangle
\end{equation}
where the $A_{\tau_i}$ are matrices (representating single 
site states), the $\langle W|, |V\rangle$ vectors (reflecting 
the influence of the
boundary conditions) acting in some auxiliary
space. One can formally define a column vector $A=(A_1,...,A_m)$ 
and rewrite (\ref{produkt})  as
\begin{equation}\label{rformal}
  |P_0\rangle_{\leftarrow} \;=\;\langle W| \,
   A^{\otimes L} |V\rangle\,\,.
\end{equation}
The existence of a nontrivial representation of (\ref{rformal}) 
can then be shown simply by defining the $A_i$'s, $\langle W|$ and 
$|V\rangle$ exactly as in ref. {\cite{sandow96}}. We use here a 
slightly different notation and define the matrix 
$A_{\tau}$ ($\tau=1,2,...,m$) by
\begin{equation}
A_{\tau}=|\tau\rangle\langle V| + \sum_{\{\tau_2\}}|\tau
\tau_2\rangle \langle \tau_2| + \sum_{\{\tau_2 \tau_3\}}|\tau
\tau_2\tau_3\rangle\langle \tau_2\tau_3|  + ...\quad .
\end{equation}
The vacuum vector is denoted by $|V\rangle$. This definition implies
\begin{equation}
A_{\tau_1}A_{\tau_2}...A_{\tau_L}|V\rangle=|\tau_1 \tau_2
... \tau_L\rangle\, .
\end{equation}
One can then choose $\langle W|$ such that (\ref{rformal}) holds by
definition. \\
Since $|p_0\rangle_{\leftarrow}$ is defined as the stationary state
one has
$T_{\leftarrow}|p_0\rangle_{\leftarrow}=|p_0\rangle_{\leftarrow}$.
The simplest mechanism to achieve this is to imagine that ${\cal R}$ 
produces a 'defect' $\bar{A}$ which is transported through the sequential 
action of ${\cal T}$  
to the left end to the chain where it disappears by applying 
${\cal  L}$. This mechanism reads then:
\begin{eqnarray}
{\cal T}[A\otimes\bar{A}] = \bar{A}\otimes A\, ,\label{bulk}\\
{\cal R}[A]|V\rangle = \bar{A}|V\rangle\, ,\label{rrand}\\
\langle W|{\cal L}[\bar{A}] = \langle W|A\, .\label{lrand}
\end{eqnarray}
By defining 
\begin{equation}\label{abar}
\bar{A}\otimes A^{\otimes^{L-1}}|V\rangle = 
\prod_{i=1}^{L-1}T_i R {A^{\otimes^{L}}|V\rangle}
\end{equation}
with 
\begin{eqnarray}
R &=& \eins\otimes ... \otimes\eins\otimes{\cal R}\,,\\
T_i &=& \eins\otimes\eins ... \otimes{\cal T}_{i}
\otimes\eins ... \otimes\eins\,
,
\end{eqnarray}
it can be shown analogously to \cite{sandow96} 
that in fact the defined ${\bar{A}}$
fulfills the algebra (\ref{bulk})-(\ref{lrand}). \\
This means that in order to solve such a model exactly, the problem of
solving the master equation has been reformulated in terms of finding
representations of the algebra.
Up to now, explicit representations of this algebra
are only known for the asymmetric exclusion model 
(\cite{hinri},\cite{wir},\cite{peschl}).\\
We are now able to write down the stationary states for the
updates $T_{\rightarrow}$ and $T_{\parallel}$:
\begin{equation}\label{lformal}
  |P_0\rangle_{\rightarrow} \;=\;\langle W| \,
   \bar{A}^{\otimes L} |V\rangle\,\, ,
\end{equation}
 and
\begin{equation}\label{pformal}
  |P_0\rangle_{\parallel} \;=\;\langle W| \,
   \bar{A}\otimes A \otimes \bar{A} 
\otimes A \otimes ... \otimes\bar{A}\otimes A|V\rangle\,\,.
\end{equation}
By applying $T_{\rightarrow}$ on (\ref{lformal}), $T_{\parallel}$ on
(\ref{pformal}) and using the algebra (\ref{bulk})-(\ref{lrand}) one
can confirm that these states are in fact the desired stationary
states. One still has to confirm that these expressions are not equal
to zero. To see this we define 
\begin{equation}
C=\sum_{i=1}^{m}A_i\quad ; \quad
\bar{C}=\sum_{i=1}^{m}\bar{A_i} \quad .
\end{equation}
The matrices ${\cal{T}}$,${\cal{R}}$,${\cal{L}}$ have the property
that their columns add up to one since the update of a local state 
has to result in some other local state. Adding up each
equation (\ref{bulk})-(\ref{lrand}) 
one obtains
\begin{eqnarray}
[C,\bar{C}] & = & 0\label{kommu}\, ,\\
\bar{C}|V\rangle & = & C|V\rangle\label{rechts}\, ,\\
\langle W|\bar{C} & = & \langle W|C\label{links}\, .
\end{eqnarray}
This implies immediately that the norm $\langle W|C^{L}|V\rangle$ of
$|p_0\rangle_{\leftarrow}$ is equal to the norms of 
$|p_0\rangle_{\rightarrow}$ and $|p_0\rangle_{\parallel}$ which shows
that the latter states are nontrivial.\\
One can now compare the densities of
the stationary states produced by the different types of update.
Using the formulas (\ref{rformal}), (\ref{lformal}), (\ref{pformal})
and (\ref{kommu})-(\ref{links})
 one gets for the densities $\rho_i(x)$, which are
defined as the probability to find a particle of type $\tau=1,...,m$
at location $i=1,...,L$ in the stationary state:
\begin{equation}\label{dichte}
{\rho_{\tau}(i)}_{\parallel}=\left\{ \begin{array}{r@{\quad:\quad}l}
    {\rho_{\tau}(i)}_{\rightarrow} &
    i \,{\mbox {odd}} \\ {\rho_{\tau}(i)}_{\leftarrow} & i\,{\mbox {even}}
    \end{array} \right.\, .
\end{equation}
Note that reversing the order of the the two time steps in
the sublattice parallel update would result 
in reverse arrows in (\ref{dichte}).\\
In order to get similar expressions for the correlation functions 
we now make use of the identity 
\begin{equation}
|V\rangle\langle V|+\sum_{l=1,2,...}\sum_{\{ \tau_1,\tau_2,...,\tau_l \} }|
\tau_1,\tau_2,...,\tau_l\rangle\langle
\tau_1,\tau_2,...,\tau_l|=\eins\, ,
\end{equation}
which expresses the completeness of our Fock space.\\
This can be rewritten as
\begin{equation}
\label{vollst}
\sum_{l=0,1,2,...}C^l|V\rangle\langle V|{(C^l)}^{\dag}=\eins\quad.
\end{equation}
One can now multiply both sides of (\ref{vollst}) with $\bar{C}$ and
make use of (\ref{kommu}) and (\ref{rechts}) which leads to
\begin{equation}
\label{c=c}
C\, =\, \bar{C}\, .
\end{equation}
This result can actually help to construct explicit representations of
the algebra since it reduces the number of matrices $2m$ which have to
be representated by one. It also reduces the $m^2$ equations for the
bulk algebra by one and the $2m$ equations for the boundary algebra
by two, which can be of significant help. \\ Note that in the case
of random sequential update the same considerations would lead to
the result ${\bar C}=0$, which is 
fulfilled by the matrix product solution of
Derrida et al of the asymmetric exclusion model \cite{derrida93}.\\
One further obtains a relationship between the
correlation functions in the stationary states which reads
for the two-point correlation function 
(generalizations to $n-$point correlation functions are obvious): 
\begin{equation}
{\langle \tau_i\tau_j\rangle}_{\parallel}\, 
=\left\{ \begin{array}{r@{\quad:\quad}l}
    {\langle \tau_i\tau_j\rangle}_{\rightarrow} &
    i,j \,{\mbox {odd}} \\ 
{\langle \tau_i\tau_j\rangle}_{\leftarrow} & i,j\,{\mbox {even}} 
\end{array} \right.
\end{equation}
The correlations between odd-even and even-odd sites cannot be
expressed so easily. Since the matrix $C$ acts like a transfer matrix
between the points of the correlation functions, the relevant length
scales are determined by $C$ {\cite{hinri}}
and will therefore be equal for all three updates.\\
Since the time evolution of a system can be written as
$|p_t\rangle =T^{t}|p_{t=0}\rangle $ it is useful to ask for the 
general eigenvectors $|p^{E}\rangle $ with 
eigenvalue $E$ of $T$. It turns out
that these can also be constructed (\cite{sandow},\cite{prep}) by
writing $E={\epsilon}^{L-1}\epsilon_{l}\epsilon_{r}$ and
multiplying the r.h.s of (\ref{abar}) with
$\frac{1}{{\epsilon}^{L-1}\epsilon_{r}}$. Details will be given in
\cite{prep}. It turns also out that by setting $\epsilon =1$
and $\epsilon_{l}=\frac{1}{\epsilon_{r}}$ one has 
the freedom to substitute (\ref{c=c}) with
${\bar{C}}=\frac{1}{\epsilon_{r}}C$.\\
In this letter we have shown that the matrix product formalism 
can be used to prove the physical equivalence of the sequential
and the sublattice parallel update. One gets explicit 
nontrivial connections
between densities and correlation functions in the stationary state.\\
One can make similar considerations concerning the relation
between the stationary states of the random sequential update
and the updates discussed in the present work. This work is in
progress \cite{prep}.\\
  \\
 \\
\noindent
{\bf Acknowledgments}\\[2mm]
This work has been performed within the research program of the 
Sonderforschungsbereich 341 (K\"oln-Aachen-J\"ulich).
We would like to thank A.~Schadschneider and  H.~Niggemann
for fruitful discussions.
 \\
 \\


\begin{thebibliography}{99}
\def\ll #1 #2 #3{{\bf #1} (19#2) #3}
\def\nm #1 #2 {\bibitem{#1} #2}
\def\gr #1 {{\large #1}}
\def\PRA{{\em Phys.\ Rev.\ }A } 
\def\PRB{{\em Phys.\ Rev.\ }B }
\def\PRE{{\em Phys.\ Rev.\ }E }
\def\PRL{{\em Phys.\ Rev.\ Lett.\ }}
\def\JPA{{\em J.\ Phys.\ A: Math\ Gen.\ }}
\def\JPC{{\em J.\ Phys.\ C: Solid State Phys.\ }}
\def\JP{{\em J.\ Physique }}
\def\JPI{{\em J.\ Phys.\ I France\ }}
\def\EPL{{\em Europhys.\ Lett.\ }}
\def\ZPB{{\em Z.\ Phys.\ }B }
\def\JSP{{\em J.\ Stat.\ Phys.\ }}
%
\nm zia {Schmittmann B and Zia R K P, {\it Statistical mechanics of
    driven diffusive System}, Academic Press, London 1995}
\nm ligget {Ligget T, {\it Interacting particle systems}, Springer,
    New York 1985}
\nm derrida93 {Derrida B, Evans M R, Hakim V and 
Pasquier V 1993 \JPA {\bf 26} 1493}
\nm traffic {Schreckenberg M, Schadschneider A, Nagel K 
and Ito M 1995 \PRE {\bf 51} 2339}
\nm hinri {Hinrichsen H 1996 \JPA {\bf 29} 3659}
\nm wir {Rajewsky N, Schadschneider A and Schreckenberg M 1996 \JPA
  {\bf 29} L305}
\nm peschl {Honecker A and Peschel I 1996 cond-mat/9606053}
\nm prep {Rajewsky N, Santen L, Schadschneider A 
and Schreckenberg M, in preparation}
\nm sandow96 {Sandow S and Krebs K 1996 cond-mat/9610029}
\nm sandow {Sandow S and Krebs K, private communication}
\end{thebibliography}
\end{document}